\documentstyle[12pt]{article}
\begin{document}

\title{ALGEBRAIC FORMULATION OF THE OPERATORIAL PERTURBATION THEORY. PART I}

\author{Ary W. Espinosa M\"uller \\
{\it Departamento de F{\'\i}sica, Universidad de Concepci\'on}\\
{\it Casilla 4009, Concepci\'on, Chile}\\ 
{\ }\\ 
Adelio R. Matamala V\'asquez \\
{\it Departamento de F{\'\i}sico--Qu{\'\i}mica, Universidad de 
Concepci\'on}\\
{\it Casilla 3--C, Concepci\'on, Chile}}

\maketitle

\begin{abstract}

A new totally algebraic formalism based on general, abstract ladder operators 
has been proposed. This approach heavily grounds in the superoperator 
formalism of Primas. However it is necessary to introduce many improvements 
in his formalism. In this regard, it has been introduced a new set of 
superoperators featured by their algebraic structure. Also, two lemmas and 
one theorem have been developed in order to algebraically reformulate the 
theory on more rigorous grounds. Finally, we have been able to build a 
coherent and self--contained formalism independent on any matricial 
representation, removing in this way the degeneracy problem.

\end{abstract}

\newpage

\section{INTRODUCTION}

The fundamental problem in perturbation theory is the solution of the
Schr\"odinger equation 

\begin{equation}
\hat{H}\Psi=E\Psi\,,
\end{equation}

for the stationary states $\Psi(x,y,z)$ of a system where the Hamiltonian 
$\hat{H}$ is split into an unperturbed Hamiltonian $\hat{H}^\circ$ and a
perturbation $\hat{V}$. Traditional treatments of the theory lean heavily on
the expansion of correction to an eigenfuction in terms of a complete set of
normalized eigenfunctions of $\hat{H}^\circ$ [1--4]. However, the problem can 
also be formulated in terms of obtaining an effective Hamiltonian 
$\hat{M}=\hat{U}\hat{H}\hat{U}^\dagger$, with $\hat{U}$ a unitary operator. 
The unitary or canonical transformation [5] method originated by Van Vleck 
[6], has been adopted by Primas [7], J\o rgensen and Pedersen [8], Mukherjee 
{\it et al.} [9] and others [10]. The $\hat{U}$ operator is unitary in the 
Van Vleck and Primas' formalism and produces a Hermitian effective 
Hamiltonian.

Murray [11] and Primas [7], have been able to show that any perturbation
theory can be formulated in the domain of the Lie algebras, in this case
generated by $\hat{H}^\circ$ and $\hat{V}$. In that concern, the solution of 
a perturbation problem is closely connected with the solution of commutator 
equations of a given type. Further, using the spectral resolution of 
$\hat{H}^\circ$, Primas was able to show that the general solution can be 
written more adequately with the aid of the superoperator algebra.

In the above scenario, our main aim is to recast the superoperator formalism
of Primas in an algebraic form using, to that end, the basic theory of ladder 
operators [12] thus our work will be reduced to prove that formally it is 
always possible to build a realization. In Part 2 of this series, we will 
show how particular realizations will lead us to successfully check the
present approach to of the perturbation theory ({\bf AFOPT}, {\em Algebraic 
Formulation of the Operator Perturbation Theory}).

The above {\bf AFOPT} avoids the matrix representation, since as it is well
known in the commonly used treatments, the perturbative series and hence the
expectation values of $\hat{H}$, depend crucially on the orthonormal
eigenbase of $\hat{H}^\circ$.

The outline of the paper is as follow. The treatment begins with the 
definition of the eigenbase $\{|n^\circ\rangle\}$ of $\hat{H}^\circ$. Then, 
the ladder operators defined in this eigenbase have been presented with their 
main characteristics. At the same time in this Sect. 2 the multilinear 
operators $\hat\eta_+^m\hat\eta_-^n$ and $\hat\eta_-^m\hat\eta_+^n$ have been 
stated. These operators will serve to establish a resolution of any operator 
belonging to the operator space ${\cal T}$, whose base has been given by 
$\{|n^\circ\rangle\}$. In Sect. 3, two lemmas and one fundamental theorem to 
of the {\bf AFOPT} are presented. In Sect. 4, the perturbation operator 
theory is briefly presented. This section is followed by a summary and 
discussions in Sect. 5 Finally, the paper ends up with the mnemonic technique 
in order to write the commutator equations.

\section{FORMALISM}

\subsection{LADDER OPERATORS}

The full Hamiltonian $\hat{H}$ is split into a zero-order Hamiltonian 
$\hat{H}^\circ$ and a perturbation $\hat{V}$ 

\begin{equation}
\begin{tabular}{ll}
$\hat{H}=\hat{H}^\circ+\lambda\,\hat{V}$&,with $\lambda\in[0,1]$
\end{tabular}
\end{equation}

Orthonormal eigenkets of $\hat{H}^\circ$ which belong to the zeroth--order
eigenspace of energy $\varepsilon_n^\circ$ are denoted by $|n^\circ\rangle$ 

\begin{equation}
\hat{H}|n^\circ\rangle=\varepsilon_n^\circ|n^\circ\rangle
\end{equation}

As the perturbation is switched on the zero-order eigenkets $|n^\circ\rangle$ 
evolves into orthonormal perturbed eigenkets $|n\rangle$ of energy 
$\varepsilon_n$.

Some time ago, De la Pe\~na and Montemayor [12--16] have shown that given the 
discrete spectral resolution of a linear and Hermitian operator $\hat{P}$, it 
is always possible to construct raising and lowering operators associated to 
that operator. Hence, related to $\hat{H}^\circ$ we have at our disposal the 
discrete eigenbase $\{|n^\circ\rangle\}$, thus we may state with all 
generality 

\begin{equation}
\hat\eta_+=\sum_nc_n|n+1\rangle \langle n|
\end{equation}

and 

\begin{equation}
\hat\eta_-=\sum_nc_{n-1}^{*}|n-1\rangle \langle n|
\end{equation}

From the orthonormality condition it is easy to see that $\hat\eta_+$ and 
$\hat\eta_-$ are ladder operators 

\begin{equation}
\hat\eta_+|k\rangle =c_k|k+1\rangle
\end{equation}

\begin{equation}
\hat\eta_-|k\rangle =c_{k-1}^{*}|k-1\rangle
\end{equation}

Now, since $\hat\eta_+$ and $\hat\eta_-$ are adjoint to each other, the 
eigenbase $\{|n\rangle\}$ is a common eigenbase to both operators 
$\hat\eta_+\hat\eta_-$ and $\hat\eta_-\hat\eta_+$ 

\begin{equation}
\hat\eta_+\hat\eta_-|n\rangle =|c_n|^2|n\rangle
\end{equation}

\begin{equation}
\hat\eta_-\hat\eta_+|n\rangle =|c_{n-1}|^2|n\rangle
\end{equation}

The coefficients $c_n$ and $c_{n-1}^*$ are complex number related to the
eigenvalues of $\hat\eta_+\hat\eta_-$ and $\hat\eta_-\hat\eta_+$.

Furthermore, we assume that the eigenvalue spectrum is bounded from below
and from above [13,17,18] 

\[
\varepsilon_0^\circ<\varepsilon _1^\circ<\cdots<\varepsilon _N^\circ 
\] 

Therefore 

\[
c_{-1}=c_M=0 
\]

From Eqs. 2.7 and 2.8 it follows that $\hat\eta_+\hat\eta_-$ differs from 
$\hat\eta_-\hat\eta_+$. In order to have only one kind of expressions, we 
adopt the {\em normal ordering}, by which the normal product of a set of 
raising and lowering operators is defined to be the product arranged, so that 
the raising operators are to the left of the lowering operators.

\subsection{SUPEROPERATORS}

Now, in order to build the algebraic formulation to of the perturbation 
theory, let us introduce the notion of {\em superoperator} [7,18,19]. The 
superoperator algebra of all linear operators acting on the wavefunction 
space ${\cal H}$, is a linear vector space, called operator space ${\cal T}$. 
Just as we define mappings $\hat{T}:{\cal H}\rightarrow{\cal H}$ called 
operators, so we can define mappings $\tau:{\cal T}\rightarrow{\cal T}$ 
called superoperators. Both kinds of mappings are linear mappings. Also, 
linearity, the sum and the product by scalar, of superoperators are defined
analogously to the definitions for the operators. Then it is clear that the
superoperator space is again a linear space. The foregoing clarification is
relevant for forthcoming developments of the theory. Actually, let us look
for the connection between operators and superoperators in the present
algebraic approach to the perturbation theory.

So as to do that, let us consider an operator $\hat{A}$ of the operator space 
${\cal T}$, we will assume that it is possible to write in normal ordering 
the following expansion 

\begin{equation}
\hat{A}=\sum_m\sum_na_{mn}\hat\eta_+^m\hat\eta_-^n
\end{equation}

Where now the $a_{mn}$ coefficients will depend on the explicit form of the
operator $\hat{A}$. It is immediate to write: 

\begin{equation}
\hat{A}=\sum_ma_{mm}\hat\eta_+^m\hat\eta_-^m+\sum_{m\neq}\sum_na_{mn}\hat\eta
_+^m\hat\eta_-^n
\end{equation}

Then it is possible to show that 

\begin{equation}
\lbrack \hat{H}^\circ,\hat\eta_+^m\hat\eta_-^m]=\hat{0}
\end{equation}

if $m=n$, and 

\begin{equation}
\lbrack \hat{H}^\circ,\hat\eta_+^m\hat\eta_-^n]\neq \hat{0}
\end{equation}

if $m\neq n$.

In fact, having in mind Eq. 2.2 and the expansion of the operator $\hat{A}$, 
we get for any ket $|k\rangle$: 

\begin{equation}
\lbrack\hat{H}^\circ,\hat\eta_+^m\hat\eta_-^n]|k\rangle=(\varepsilon_{k+m-n}^ 
\circ-\varepsilon_k^\circ)\hat\eta_+^m\hat\eta_-^n|k\rangle
\end{equation}

from which the results Eq. 2.11 and Eq. 2.12 follow.

Then it is feasible to define the following operators 

\begin{equation}
\hat{A}_\parallel=\sum_ma_{mm}\hat\eta_+^m\hat\eta_-^m
\end{equation}

and 

\begin{equation}
\hat{A}_\perp=\sum_{m\neq}\sum_n a_{mn}\hat\eta_+^m\hat\eta_-^n
\end{equation}

Therefore 

\begin{equation}
\hat{A}=\hat{A}_\parallel+\hat{A}_\perp
\end{equation}

The operators $\hat{A}_\parallel$ and $\hat{A}_\perp$ are referred to as the 
{\em parallel} and {\em orthogonal} components of the operator $\hat{A}$ 
relative to $\hat{H}^\circ$. They satisfy the next relations: 

\begin{equation}
\lbrack\hat{H}^\circ,\hat{A}_\parallel]=\hat{0}
\end{equation}

and 

\begin{equation}
\lbrack\hat{H}^\circ,\hat{A}_\perp]\neq \hat{0}
\end{equation}

Since $\hat{A}$ is any operator belonging to space ${\cal T}$, we have split 
the operator space ${\cal T}$ into two subspaces ${\cal T}_\parallel$ and 
${\cal T}_\perp$. Where ${\cal T}_\parallel$ contains all the operators that 
commute with $\hat{H}^\circ$, and ${\cal T}_\perp$ all the operators that do 
not commute with $\hat{H}^\circ$. It is necessary to remark that 

\begin{equation}
{\cal T}_\parallel\cup {\cal T}_\perp={\cal T}
\end{equation}

and 

\begin{equation}
{\cal T}_\parallel\cap {\cal T}_\perp=\{\hat{0}\}
\end{equation}

As it has been pointed out, the operator space ${\cal T}$ is a vector space, 
therefore Eq. 2.16 may be interpreted as the resolution of operator $\hat{A}$ 
into two components: one parallel component relative to $\hat{H}^\circ$ and 
other orthogonal component relative to $\hat{H}^\circ$. The above remark 
contains the key which will lead us to prove the theorem about the existence 
and uniqueness of the inverse of a superoperator $\Gamma$ ( see Sect. 3 ). 
The partitioning that has been performed is equivalent to the partitioning in 
block-diagonal and off-diagonal of Primas [7], and this in turn is the same 
partitioning as the even and odd one of J\o rgensen and Pedersen [8].

\section{TWO LEMMAS AND ONE THEOREM}

As was distinguished by Murray [20] and by Primas [7], the solution of a
perturbation problem may be formulated in terms of the solution of the
commutator equation of the type 

\begin{equation}
\lbrack \hat{H}^\circ,\hat{X}]=\hat{Y}
\end{equation}

where $\hat{H}^\circ$ is the unperturbed Hamiltonian, $\hat{Y}$ an operator 
or function of operators and $\hat{X}$ is an unknown operator that has to be 
determined. Using the spectral resolution of $\hat{H}^\circ$, Primas [7] has 
been able to state the general solution for Eq. 3.1 in the language of 
superoperator, as given by 

\begin{equation}
\hat{X}-\Pi (\hat{X})=\Gamma ^{-1}(\hat{Y})
\end{equation}

In Eq. 3.2 $\Pi$ represents the superoperator that {\em projects from any
operator, that part which} {\em commutes with} $\hat{H}^\circ$, and 
$\Gamma^{-1}$ denotes the inverse of the superoperator $\Gamma$ called {\em 
derivation superoperator generated by} $\hat{H}^\circ$ [7]. Our task will be 
to reformulate Eq. 3.2 in the abstract ladder operator language. If we are 
able to represent the $\Pi$, $\Gamma$ and $\Gamma^{-1}$ superoperators in 
terms of the abstract $\hat\eta_+$ and $\hat\eta_-$ ladder operators of the 
Sect. 2, we will have achieved the main goal of the present work. To do that, 
we would like to state two lemmas. Before doing that, we will define 
$\Pi(\hat{X})$ as {\em the parallel projection} of the $\hat{X}$ operator. 
\footnote{$\Pi(\hat{X})$, $\Gamma(\hat{X})$ and $\Gamma^{-1}(\hat{X})$ in our 
notation correspond to $\langle\hat{X}\rangle$, $k(\hat{X})$ and $\frac 
1k(\hat{X})$ in that of Primas [7].}

{\bf Definition:} {\em For any linear and Hermitian operator} 
$\hat{X}\in{\cal T}$ {\em the parallel projection will be defined by} 

\begin{equation}
\Pi(\hat{X})=\sum_n\langle n^\circ|\hat{X}|n^\circ\rangle|n^\circ\rangle 
\langle n^\circ|
\end{equation}

{\bf Lemma 1:} {\em Given the abstract ladder operators} $\hat\eta_+$ {\em 
and} $\hat\eta_-$ {\em the parallel projection superoperator defined over the 
multilinear operators} $\hat\eta_+^m\hat\eta_-^n$, $m,n=0,1,2,\cdots$ {\em 
satisfies the following relation} 

\begin{equation}
\Pi(\hat\eta_+^m\hat\eta_-^n)=\delta_{mn}\hat\eta_+^m\hat\eta_-^n
\end{equation}

{\bf Proof:} The action of the multilinear operator 
$\hat\eta_+^m\hat\eta_-^n$ on any ket $|k\rangle$ may be represented by 

\[
\hat\eta_+^m\hat\eta_-^n|k\rangle=\lambda(k;m,n)|k+m-n\rangle 
\] 

where $\lambda(k;m,n)$ is a multiplicative factor depending on the powers $m$ 
and $n$ and the quantum number $k$. By definition 

\[
\Pi(\hat\eta_+^m\hat\eta_-^n)=\sum_k\langle k|\hat\eta_+^m\hat\eta_-^n|k 
\rangle|k\rangle\langle k|
\]

and rearranging 

\[
\Pi(\hat\eta_+^m\hat\eta_-^n)=\sum_k\lambda(k;m,n)\langle k|k+m-n\rangle|k 
\rangle\langle k|
\]

\[
\Pi(\hat\eta_+^m\hat\eta_-^n)=\sum_k\lambda(k;m,n)\delta_{mn}|k\rangle\langle 
k|
\]

\[
\Pi(\hat\eta_+^m\hat\eta_-^n)=\delta_{mn}\sum_k\lambda(k;m,n)|k\rangle\langle 
k|
\]

\[
\Pi(\hat\eta_+^m\hat\eta_-^n)=\delta_{mn}\hat\eta_+^m\hat\eta_-^n 
\] 

which proves Lemma 1

The next property derives from the definiton of $\Pi$ itself : 

\begin{equation}
\Pi (\alpha \hat{A}+\beta \hat{B})=\alpha \Pi (\hat{A})+\beta \Pi (\hat{B})
\end{equation}

From Eq. 3.5 and Lemma 1 it is easy to obtain the properties 

\begin{equation}
\Pi (\hat{A})=\hat{A}_\parallel
\end{equation}

\begin{equation}
\Pi (\hat{A}_\parallel)=\hat{A}_\parallel
\end{equation}

\begin{equation}
\Pi (\hat{A}_\perp)=\hat{0}
\end{equation}

Furthermore, from Eqs.2.16 and 3.4 we may deduce the useful identity 

\begin{equation}
\hat{A}_\perp=\hat{A}-\Pi (\hat{A})
\end{equation}

{\bf Definition :} The {\em derivation superoperator} $\Gamma$ is given by 

\begin{equation}
\Gamma(\hat{X})=[\hat{H}^\circ,\hat{X}]
\end{equation}

with $\hat{X}\in{\cal T}$.

To study this superoperator, it is necessary to state the following lemma.

{\bf Lemma 2:} {\em Given the operator} $\hat{H}^\circ$ {\em and its ladder 
operators} $\hat\eta_+$ {\em and} $\hat\eta_-$ {\em the derivation 
superoperator of the multilinear operator} $\hat\eta_+^m\hat\eta_-^n\in{\cal 
T}$ {\em satisfies the following general form}: 

\begin{equation}
\Gamma(\hat\eta_+^m\hat\eta_-^n)=\hat\eta_+^m\hat\eta_-^n\sum_k(\varepsilon_{
k+m-n}^\circ-\varepsilon_k^\circ)|k\rangle\langle k|
\end{equation}

{\bf Proof:} By definition of $\Gamma$ we get 

\begin{equation}
\Gamma(\hat\eta_+^m\hat\eta_-^n)|k\rangle=\varepsilon_{k+m-n}^\circ\hat\eta_+ 
^m\hat\eta_-^n|k\rangle-\varepsilon_k^\circ\hat\eta_+^m\hat\eta_-^n|k\rangle
\end{equation}

Multiplying to the right by the bra $\langle k|$ and summing up, it follows 

\[
\Gamma(\hat\eta_+^m\hat\eta_-^n)=\sum_k(\varepsilon_{k+m-n}^\circ-\varepsilon 
_k^\circ)\hat\eta_+^m\hat\eta_-^n|k\rangle\langle k|
\]

From which Lemma 2 has been proved.

The next properties are easily derived from the definition of the $\Gamma$ 
superoperator.

Since $\Gamma$ is a linear superoperator one has 

\begin{equation}
\Gamma(\alpha\hat{A}+\beta\hat{B})=\alpha\Gamma(\hat{A})+\beta\Gamma(\hat{B})
\end{equation}

Also, it is immediate that 

\begin{equation}
\Gamma (\hat{A}_\parallel)=\hat{0}
\end{equation}

\begin{equation}
\Gamma (\hat{A}_\perp)\neq \hat{0}
\end{equation}

and since $\Gamma$ is the superoperator which forms the commutator from any
operator of ${\cal T}$ with $\hat{H}^\circ$, one gets 

\begin{equation}
\Gamma (\hat{A}\hat{B})=\hat{A}\Gamma (\hat{B})+\Gamma (\hat{A})\hat{B}
\end{equation}

The superoperator $\Gamma$ obtains its name from its derivative properties.

Some comments must be deserved to the last two lemmas. Firstly, from Eq. 3.4 
one realizes that {\em the action of} $\Pi$ {\em is independent on the 
physics of the system, since the Hamiltonian has not been considered 
explicitly}. Hence the superoperator $\Pi$ simply split the entire operator
space into two subspaces (orthogonal and parallel). Secondly, Eq. 3.6 points 
out directly, that {\em the action of} $\Gamma$ {\em has an explicit 
dependence on} $\hat{H}^\circ$, due to the presence of the transition energy 
$\Delta\varepsilon^\circ=\varepsilon_{k+m-n}^\circ-\varepsilon_k^\circ$, 
which is also an immediate consequence of the definition of $\Gamma$ itself.

One very fundamental question to build a coherent and self contained 
algebraic perturbation theory, is to assure the existence of the 
superoperator $\Gamma^{-1}$ in the Primas' theory. Primas has prevented from 
demostrating this relevant theorem because he considers that the inverse 
superoperator $\Gamma^{-1}$ has the whole operator space ${\cal T}$ as its 
domain [7]. On the contrary, we will show that $\Gamma^{-1}$ exists solely in 
the orthogonal subspace ${\cal T}_\perp{\cal\subset T}$. Therefore, we aim to 
discover the proper arguments leading to demostrate the existence and 
uniqueness of inverse superoperator. A subject that we will now study in 
somewhat greater detail.

{\bf THEOREM:} {\em The inverse superoperator} $\Gamma^{-1}$ {\em exists and 
it is unique, if and only if the domain and the range of the linear mapping 
associated with it, can be adequately restricted to the orthogonal subspace} 
${\cal T}_\perp{\cal\subset T}$.

{\bf Proof:} Since the superoperator $\Gamma$ is a linear mapping, it allows 
us to introduce the {\em kernel of a linear mapping} [21] and hence the 
kernel of the superoperator $\Gamma$, which we denote by $\ker\Gamma$, and 
that we define as the set of all the operators $\hat{X}\in{\cal T}$ such that 
$\Gamma(\hat{X})=\hat{0}$.

Having in mind that a linear mapping whose kernel is $\{\hat{0}\}$, is 
injective [21,22], we find that $\Gamma$,defined by 

\begin{equation}
\Gamma :{\cal T}\rightarrow {\cal T}
\end{equation}

with 

\[
\Gamma (\hat{X})=[\hat{H}^\circ,\hat{X}] 
\]

is not an injective mapping. Really, Eqs 3.14 and 3.15 show that 
$\ker\Gamma={\cal T}_\parallel\neq\{\hat{0}\}$. However, it is possible to
redefine the domain and the range of the mapping $\Gamma$ to the orthogonal
subspace, since $\Pi(\Gamma(\hat{X}))=\hat{0}$. Thus redefining the mapping 
$\Gamma$ by : 

\begin{equation}
\Gamma :{\cal T}_\perp\rightarrow {\cal T}_\perp
\end{equation}

with 

\[
\Gamma (\hat{X})=[\hat{H}^\circ,\hat{X}] 
\]

we succeed in getting $\ker\Gamma=\{\hat{0}\}$.

Actually, if we assume that an arbitrary orthogonal operator, 
$\hat{A}\in{\cal T}_\perp$, is such that $\hat{A}\in\ker\Gamma $, then 
$\Gamma(\hat{A})=\hat{0}$. But, we know that $\Gamma(\hat{A})\neq\hat{0}$ if 
$\hat{A}\in{\cal T}_\perp$, then the assumption is false. Hence the unique 
element of the $\ker\Gamma$ is $\hat{0}$. In other words, $\Gamma$ is 
injective. Otherwise, the image and the range of $\Gamma$ are the same, so 
$\Gamma$ must be surjective. Therefore, the inverse of the $\Gamma$ exists 
and is unique. Hence, by fair means we can now write 

\begin{equation}
\Gamma ^{-1}(\Gamma (\hat{X}))=\Gamma (\Gamma ^{-1}(\hat{X}))=\hat{X}
\end{equation}

if and only if 

\begin{equation}
\hat{X}\in {\cal T}_\perp
\end{equation}

and the Theorem has been proved.

Lastly the following properties are evident from $\Gamma^{-1}$,since the
linearity of $\Gamma^{-1}$ follows from the linearity of $\Gamma$, 

\begin{equation}
\Gamma^{-1}(\alpha\hat{A}+\beta\hat{B})=\alpha\Gamma^{-1}(\hat{A})+\beta
\Gamma^{-1}(\hat{B})
\end{equation}

Thus the perturbational problem has been reduced to the finding of an 
explicit expression for $\Gamma^{-1}$. In Part 2 of this series, we will 
study particular forms for $\Gamma^{-1}$ (also for $\Gamma$ and $\Pi$), 
depending on the algebra of ladder operators associated to the physical 
problem to be tackled.

\section{PERTURBATION METHOD}

As aforementioned the complete Hamiltonian $\hat{H}$ has been split into an 
unperturbed Hamiltonian $\hat{H}^\circ$ and a perturbation operator $\hat{V}$ 
scaling with the real parameter $\lambda\in[0,1]$ 

\begin{equation}
\hat{H}=\hat{H}^\circ+\lambda \hat{V}
\end{equation}

Besides, the comments that have been made at the begining of Sect. 2 (cf. 
Eqs. 2.1 and 2.2) also special mention deserves the fact that in general 

\begin{equation}
\lbrack \hat{H}^\circ,\hat{V}]\neq \hat{0}
\end{equation}

which implies that we cannot find a common eigenbase for $\hat{H}^\circ$ and 
$\hat{V}$. But we can think of a certain unitary transformation, that will 
change this situation.

The idea of choosing a unitary transformation corresponds to the need of 
leaving invariant the spectrum of eigenvalues of the energy. The unitary 
transformation only modifies the eigenvectors.

Let $\hat{U}$ be a unitary transformation defined as 

\begin{equation}
\hat{U}\hat{H}\hat{U}^\dagger=\hat{U}(\hat{H}^\circ+\lambda\hat{V})\hat{U}^
\dagger
\end{equation}

We can now introduce two new operators $\hat{M}$ and $\hat{W}$, through the
definitions 

\begin{equation}
\hat{M}=\hat{U}\hat{H}\hat{U}^\dagger
\end{equation}

and 

\begin{equation}
\hat{W}=\hat{M}-\hat{H}^\circ
\end{equation}

The relation 4.4 allows to write 

\begin{equation}
\hat{M}=\hat{H}^\circ+\hat{W}
\end{equation}

From Eq. 4.4 it is immediate to see that $\hat{M}$ has the same spectrum of
eigenvalues as the Hamiltonian $\hat{H}$.

We will now suppose that $\hat{U}$ satisfies the following condition 

\begin{equation}
\lbrack \hat{H}^\circ,\hat{W}]=\hat{0}
\end{equation}

That means that $\hat{H}^\circ$ and $\hat{M}$ will have common eigenvectors 
as follows from Eq. 4.6. Therefore, if Eq. 4.7 holds, we may write 

\begin{equation}
\langle n^\circ|\hat{M}|n^\circ\rangle=\langle n^\circ|\hat{H}^\circ|n^\circ 
\rangle+\langle n^\circ|\hat{W}|n^\circ\rangle
\end{equation}

\begin{equation}
\varepsilon_n=\varepsilon_n^\circ+\langle n^\circ|\hat{W}|n^\circ \rangle
\end{equation}

Since $\langle n^\circ|\hat{M}|n^\circ\rangle=\varepsilon_n$ and 
$\hat{M}=\hat{U}\hat{H}\hat{U}^\dagger$ we may write 

\begin{equation}
\hat{U}\hat{H}\hat{U}^\dagger|n^\circ\rangle=\varepsilon_n|n^\circ\rangle
\end{equation}

Therefore, after multiplying to the left by $\hat{U}^\dagger$ and having in 
mind that $\hat{U}$ is a unitary transformation 

\begin{equation}
\hat{H}\hat{U}^\dagger|n^\circ\rangle=\varepsilon_n\hat{U}^\dagger|n^\circ
\rangle
\end{equation}

where $\hat{U}^\dagger|n^\circ\rangle$ is the new eigenket of $\hat{H}$.

Briefly, imposing the condition given by Eq. 4.7 we have the following 
scheme: 

\begin{eqnarray}
&&\fbox{$|n\rangle =\hat{U}^{\dagger }|n^\circ\rangle $} \\
&&\fbox{$\varepsilon_n=\varepsilon_n^\circ+\langle n^\circ|\hat{W}|n^\circ 
\rangle$}\nonumber
\end{eqnarray}

That is to say, resolving the eigenvalue problem for the Hamiltonian 
$\hat{H}$ implies to find the transformation $\hat{U}^\dagger$ that makes 
possible the Eq. 4.7 which in turns, will allow us to write the explicit form 
of $\hat{W}$.

Let us suppose now that the unitary transformation may be written as the
exponential of a certain antihermitian operator, $\hat{G}=-\hat{G}^\dagger$, 
henceforth referred to as the {\em generator of the transformation}. Then we 
immediately get, the relation 

\[
\hat{W}=\hat{M}-\hat{H}^\circ 
\]

\[
\hat{W}=\hat{U}\hat{H}\hat{U}^\dagger-\hat{H}^\circ 
\]

\begin{equation}
\hat{W}=\exp(\hat{G})\hat{H}\exp(-\hat{G})-\hat{H}^\circ
\end{equation}

Using the expansion of Baker-Camppell-Hausdorff [23] we get 

\begin{equation}
\hat{W}=\left(\hat{H}+\frac 1{1!}[\hat{G},\hat{H}]+\frac 1{2!}[\hat{G},[\hat{ 
G},\hat{H}]]+\cdots\right) -\hat{H}^\circ
\end{equation}

From Eq. 4.1 we arrive at 

\begin{equation}
\hat{W}=\lambda\hat{V}+\frac 1{1!}[\hat{G},\hat{H}^\circ+\lambda \hat{V}]+ 
\frac 1{2!}[\hat{G},[\hat{G},\hat{H}^\circ+\lambda \hat{V}]]+\cdots
\end{equation}

Let us now assume that 

\begin{equation}
\hat{W}=\lambda\hat{W}_1+\lambda^2\hat{W}_2+\cdots
\end{equation}

and 

\begin{equation}
\hat{G}=\lambda\hat{G}_1+\lambda^2\hat{G}_2+\cdots
\end{equation}

Insertion of Eq. 4.15 and 4.16. in Eq. 4.13, furthermore, developing, 
rearranging and comparing equal powers in $\lambda$, lead us in a 
straighforward way to 

\begin{equation}
\lbrack\hat{H}^\circ,\hat{G}_1]=\hat{V}-\hat{W}_1
\end{equation}

\begin{equation}
\lbrack\hat{H}^\circ,\hat{G}_2]=\frac 1{1!}[\hat{G}_1,\hat{V}]+\frac1{2!} 
[\hat{G}_1,[\hat{G}_1,\hat{H}^\circ]]-\hat{W}_2
\end{equation}

\begin{eqnarray}
\lbrack\hat{H}^\circ,\hat{G}_3]&=&\frac 1{1!}[\hat{G}_2,\hat{V}]+\frac 1{2!}[ 
\hat{G}_1,[\hat{G}_2,\hat{H}^\circ]]+\frac 1{2!}[\hat{G}_2,[\hat{G}_1,\hat 
{H}^\circ]]+\\
&&\frac 1{2!}[\hat{G}_1,[\hat{G}_1,\hat{V}^\circ]]+\frac 1{3!}[\hat{G}_1,[
\hat{G}_1,[\hat{G}_1,\hat{H}^\circ]]]-\hat{W}_3  \nonumber
\end{eqnarray}

... an so on.

It is apparent that the set of last Eqs. 4.18-4.20 is a {\em system of 
coupled commutator equations for the} $\hat{G}_n$ {\em operators}. This set
obeys the general structure 

\begin{equation}
\lbrack\hat{H}^\circ,\hat{G}_n]=\hat{A}_n-\hat{W}_n
\end{equation}

where $\hat{H}^\circ$ and $\hat{A}_1=\hat{V}$, constitute the data of the 
problem and the $\hat{G}_n$ are the unknown operators to be determined. The 
$\hat{A}_n$ operators, with $n\neq1$, are specified in terms of 
$\hat{H}^\circ$ and $\hat{A}_m$ with $m<n$.

It is necessary to determine the $\hat{W}$ operator, provided that 
$[\hat{H}^\circ,\hat{W}]=\hat{0}$ or equivalently to that of 
$\Pi(\hat{W})=\hat{W}$. However, these conditions are fulfilled if, in turn 
each one of $\hat{W}_n$ results to be a parallel component operator relative 
to $\hat{H}^\circ$. On this basis it may be concluded that 

\begin{equation}
\Pi(\hat{W}_n)=\hat{W}_n
\end{equation}

Now the operation with $\Pi$ on Eq. 4.21 leads to 

\begin{equation}
\Pi([\hat{H}^\circ,\hat{G}_n])=\Pi(\hat{A}_n)-\Pi(\hat{W}_n)
\end{equation}

Having in mind the identity 

\begin{equation}
\Pi([\hat{H}^\circ,\hat{G}_n])=\hat{0}
\end{equation}

we get 

\begin{equation}
\Pi(\hat{A}_n)=\Pi(\hat{W}_n)
\end{equation}

Thus from Eq. 4.22 we write 

\begin{equation}
\hat{W}_n=\Pi(\hat{A}_n)
\end{equation}

Otherwise, from the definition 2 we have that 

\begin{equation}
\Gamma(\hat{G}_n)=[\hat{H}^\circ,\hat{G}_n]
\end{equation}

provided that $\hat{G}_n\in{\cal T}_\perp$, for every $n$. However, this 
condition is equivalent to say that 

\begin{equation}
\Pi(\hat{G}_n)=\hat{0}
\end{equation}

Therefore from Eq. 4.22 we obtain 

\begin{equation}
\Gamma(\hat{G}_n)=\hat{A}_n-\hat{W}_n
\end{equation}

or 

\begin{equation}
\Gamma(\hat{G}_n)=\hat{A}_n-\Pi(\hat{A}_n)
\end{equation}

But the hand right side of the above equation is an operator that belongs to 
${\cal T}_\perp$, therefore $\Gamma$ is well-defined. Thus, it may be deduced 
that $\Gamma^{-1}$ exists, in brief 

\begin{equation}
\Gamma^{-1}(\Gamma(\hat{G}_n))=\Gamma^{-1}(\hat{A}_n-\Pi(\hat{A}_n))
\end{equation}

or 

\begin{equation}
\hat{G}_n=\Gamma^{-1}(\hat{A}_n-\Pi(\hat{A}_n))
\end{equation}

To sum up, given a problem of the type 

\begin{equation}
\hat{H}=\hat{H}^\circ+\hat{V}
\end{equation}

we will have that 

\begin{equation}
\fbox{$\varepsilon_n=\varepsilon_n^\circ+\langle n^\circ|\hat{W}|n^\circ 
\rangle$}
\end{equation}

and 

\begin{equation}
\fbox{$|n\rangle=\hat{U}^\dagger|n^\circ\rangle$}
\end{equation}

Where 

\begin{equation}
\hat{W}=\lambda\hat{W}_1+\lambda^2\hat{W}_2+\cdots
\end{equation}

\begin{equation}
\hat{W}_n=\Pi(\hat{A}_n)
\end{equation}

and 

\begin{equation}
\hat{G}=\lambda\hat{G}_1+\lambda^2\hat{G}_2+\cdots
\end{equation}

\begin{equation}
\hat{G}_n=\Gamma^{-1}(\hat{A}_n-\Pi(\hat{A}_n))
\end{equation}

The explicit forms of any $\hat{A}_n$ are : 

\begin{equation}
\hat{A}_1=\hat{V}
\end{equation}

\begin{equation}
\hat{A}_2=\frac 1{1!}[\hat{G}_1,\hat{V}]+\frac 1{2!}[\hat{G}_1,[\hat{G}_1,
\hat{H}^\circ]]
\end{equation}

\begin{eqnarray}
\hat{A}_3&=&\frac 1{1!}[\hat{G}_2,\hat{V}]+\frac 1{2!}[\hat{G}_1,[\hat{G}_2,
\hat{H}^\circ]]+\frac 1{2!}[\hat{G}_2,[\hat{G}_1,\hat{H}^\circ]]\\
&&+\frac 1{2!}[\hat{G}_1,[\hat{G}_1,\hat{V}^\circ]]+\frac 1{3!}[\hat{G}_1,[
\hat{G}_1,[\hat{G}_1,\hat{H}^\circ]]]\nonumber
\end{eqnarray}

... an so on.

In order to know all the terms of the series, we have developed a mnemonic
method (Cf. appendix): 

\[
\hat{A}_1=(1) 
\]

\[
\hat{A}_2=(1|1)\oplus (1,1|0) 
\]

\[
\hat{A}_3=(2|1)\oplus(1,2|0)\oplus(2,1|0)\oplus(1,1|1)\oplus(1,1,1|0) 
\]

\section{SUMMARY AND DISCUSSIONS}

It has been shown that from the spectral resolution of $\hat{H}^\circ$, the 
abstract ladder operators $\hat\eta_+$ and $\hat\eta_-$ may be defined. In 
turn, these operators serve to build multilinear operators in normal ordering 
$\hat\eta_+^m\hat\eta_-^n$. Taking advantage of the properties of the 
$\hat\eta_+^m\hat\eta_-^n$ in relation to $\hat{H}^\circ$, we have been able 
to split the entire space ${\cal T}$ into two subspaces ${\cal T}_\parallel$ 
and ${\cal T}_\perp$ accordingly to any operator that commutes or not with 
$\hat{H}^\circ$. The above splitting of ${\cal T}$ has allowed us to 
demonstrate the existence and uniqueness of $\Gamma^{-1}$ under the condition 
that the domain and the range of $\Gamma$ must be the orthogonal subspace 
${\cal T}_\perp\subset{\cal T}$. Primas [7] was prevented from demostrating 
this relevant theorem, because he had considered that the superoperator 
$\Gamma^{-1}$ has the whole operator space ${\cal T}$ as its domain.

As may be seen from Sect. 4, the entire algebraic formulation of the operator 
perturbation method lean heavily on the well-defined 
$\Pi(\hat{\eta}_+^m\hat\eta_-^n)$, $\Gamma(\hat\eta_+^m\hat\eta_-^n)$ and 
$\Gamma^{-1}(\hat\eta_+^m\hat\eta_-^n)$ operators.

As was remarked at the begining, the present approach has been built 
independently on whatever matricial representation. Therefore, the 
Hamiltonian $\hat{H}^\circ$ may have any degeneracy, however this situation 
is immaterial in that concern the purely algebraic relations between the 
operators involved.

In Part 2 of this series, the method is seccesfully applied to two quantum
mechanical systems: ``The Stark Effect in the Harmonic Oscillator'' and ``The 
Generalized Zeeman Effect''.

\section{APPENDIX}

In order to write out efficiently the explicit form of the commutator 
equations determining the $\hat{A}_n$ operators, we have developed a mnemonic 
method.

RULE 1: A bracket of two sides is drawn 

\[
(\cdots|\cdots) 
\]

RULE 2: In the right side we must put 1 or 0.

RULE 3: In left side of the bracket we must put integers, in such way that 
its sum must be $n$, i.e. the order of the iteration, consequently the 
subindex of $\hat{A}_n$ superoperator.

RULE 4: We return to rule 1 until exhausting the possibilities of generating 
further diagrams.

RULE 5: In order to write an explicit commutator form for each operator 
$\hat{A}_n$, we must consider 

\[
\begin{tabular}{ll}
Left Side & Right Side \\ 
$1\rightarrow \hat{G}_1$ & $0\rightarrow \hat{H}^\circ$ \\ 
$2\rightarrow \hat{G}_2$ & $1\rightarrow \hat{V}$ \\ 
$3\rightarrow \hat{G}_3$ &  \\ 
$\cdots$ & 
\end{tabular}
\]

Besides, we have to remember that each expression is divided by the factorial 
of the number of integers in left side.

As an example we calculate $\hat{A}_2$ and $\hat{A}_3$:

\[
\hat{A}_2=(1|1)\oplus (1,1|0) 
\]

\[
\hat{A}_2=\frac 1{1!}[\hat{G}_1,\hat{V}]+\frac 1{2!}[\hat{G}_1,[\hat{G}_1,
\hat{H}^\circ]] 
\]

\[
\hat{A}_3=(2|1)\oplus (1,2|0)\oplus (2,1|0)\oplus (1,1|
1)\oplus (1,1,1|0) 
\]

\begin{eqnarray*}
&\hat{A}_3=\frac 1{1!}[\hat{G}_2,\hat{V}]+\frac 1{2!}[\hat{G}_1,[\hat{G}_2,
\hat{H}^\circ]]+\frac 1{2!}[\hat{G}_2,[\hat{G}_1,\hat{H}^\circ]]+& \\
&\frac 1{2!}[\hat{G}_1,[\hat{G}_1,\hat{V}^\circ]]+\frac 1{3!}[\hat{G}_1,[
\hat{G}_1,[\hat{G}_1,\hat{H}^\circ]]]&
\end{eqnarray*}

In what follows we display some diagrams: 

\[
\hat{A}_1=(1) 
\]

\[
\hat{A}_2=(1|1)\oplus (1,1|0) 
\]

\[
\hat{A}_3=(2|1)\oplus(1,2|0)\oplus(2,1|0)\oplus(1,1|1)\oplus(1,1,1|0) 
\]

\[
\hat{A}_4=(3|1)\oplus(1,3|0)\oplus(3,1|0)\oplus(2,2|0)\oplus(1,2|1)\oplus(2,1
|1)\oplus 
\]

\[
(1,1,2|0)\oplus(1,2,1|0)\oplus(2,1,1|0)\oplus(1,1,1|1)\oplus (1,1,1,1|0) 
\]

\section{ACKNOWLEDGMENTS}

We thank to Miss Paula J. Espinosa M. and Mrs. A. Hasb\'{u}n for subsequent 
helps and for reading the manuscript.

One of us (A.W.E.M.) is grateful for finantial support under FONDECYT grants
1989-0657.


\section{REFERENCES}

\begin{thebibliography}{XX}

\bibitem{[1]} E. Schr\"odinger, Ann. Phys. {\bf 80}, 437 (1926).

\bibitem{[2]} P. O. L\"owdin, {\it Perturbation Theory and its Applications 
in Quantum Mechanics}, ed. by C. H. Wilcox (Wiley, 1966).

\bibitem{[3]} P. O. L\"owdin, J. Math. Phys. {\bf 3}, 969 (1962); Adv. Phys. 
{\bf 5}, 1 (1956).

\bibitem{[4]} J. O. Hirschfelder, Int. J. Quantum Chem. {\bf 3}, 731 (1969).

\bibitem{[5]} P. A. M. Dirac, {\it The Principles of Quantum Mechanics} (
Clarendon Press, Oxford, 1959).

\bibitem{[6]} J. H. Van Vleck, Phys. Rev. {\bf 33}, 467 (1929); O. M. Jordal 
Phys. Rev. {\bf 45}, 87 (1934).

\bibitem{[7]} H. Primas, Rev. Mod. Phys. {\bf 35}, 710 (1963); Helv. Phys. 
Acta {\bf 34}, 331 (1961).

\bibitem{[8]} F. J\o rgensen, Mol. Phys. {\bf 29}, 1137 (1975); F. 
J\o rgensen and T. Pedersen, Mol. Phys. {\bf 27}, 33 (1974); {\bf 27}, 959 
(1974).

\bibitem{[9]} R. K. Moitra, D. Mukherjee and A. M. Pramana {\bf 9}, 545 
(1977); Mol. Phys. {\bf 30}, 1961 (1975); {\bf 33}, 953 (1977).

\bibitem{[10]} P. Westhans, E. G. Bradford and D. Hall, J. Chem. Phys. {\bf 
62}, 1607 (1975); P. Westhaus, Int. J. Quantum Chem. {\bf 20}, 1243 (1981).

\bibitem{[11]} T. Kato, Prog. Theor. Phys. {\bf 4}, 514 (1959); C. Bloch and 
J. Horowitz, Nucl. Phys. {\bf 8}, 91 (1958); B. H. Brandow, Rev. Mod. Phys. 
{\bf 39}, 771 (1967). 

\bibitem{[12]} L. De la Pe\~{n}a and R. Montemayor, Am. J. Phys. {\bf 48}, 
855 (1980).

\bibitem{[13]} F.M. Fern\'{a}ndez and E. A. Castro, Am. J. Phys. {\bf 52}, 
344 (1984). 

\bibitem{[14]} J. Cizek and J. Paldus, Int. J. Quantum Chem. {\bf 12}, 875 
(1977).

\bibitem{[15]} M. Berrondo and A. Palma, J. Phys. A: Math. Gen. {\bf 13}, 773 
(1980).

\bibitem{[16]} J. Morales, J. L\'opez-Bonilla and A. Palma, J. Math. Phys. 
{\bf 28}, 1032 (1987).

\bibitem{[17]} N. W. Bazley and D. W. Fox, Rev. Mod. Phys. {\bf 35}, 712 
(1963).

\bibitem{[18]} P. O. L\"owdin, Int. J. Quantum Chem. {\bf 16}, 485 (1982). 

\bibitem{[19]} J. A. Crawford, Nuovo Cimento {\bf 10}, 698 (1958); M. 
Rosenblum, Duke Math. J. {\bf 23}, 263 (1956).

\bibitem{[20]} F. J. Murray, J. Math. Phys. {\bf 3}, 451 (1962).

\bibitem{[21]} S. Lang, {\it Linear Algebra} (Wesley, 1971).

\bibitem{[22]} M. Schechter, {\it Operatorial Methods in Quantum Mechanics} 
(North Holland, 1981).

\bibitem{[23]} F. Hausdorff, Leipziger Ber. Ges. Wiss. Math. Phys. kl. {\bf 
58}, 19 (1906).

\end{thebibliography}
\end{document}